\documentclass[fleqn,twocolumn]{revtex4-1}  
\usepackage{bm,amsthm}
\usepackage{amsmath,amssymb}
\usepackage[dvips]{graphicx} 
\usepackage{tabls}

\def\ga{\alpha}

\def\gd{\delta}

\def\gp{\pi}

\def\gS{\Sigma}

\def\gl{\lambda}

\def\delrl1{\stackrel {\leftrightarrow} {\partial_1}}

\def\part{\partial}

\def\A0{A^{+}_0}

\def\xpl{x^{+}}

\def\xmin{x^{-}}

\def\ulix{\underline{x}}
\def\uliy{\underline{y}}

\newcommand{\nc}{\newcommand}
\nc{\intl}{\int\limits_{-L}^{+L}\!\!\frac{{\rm d}x^-}{2}}
\nc{\intly}{\int\limits_{-L}^{+L}\!\!{{{\rm d}y^-}\over\!2}}
\nc{\intlz}{\int\limits_{-L}^{+L}\!\!{{{\rm d}z^-}\over\!2}}
\nc{\intlu}{\int\limits_{-L}^{+L}\!\!{{{\rm d}u^-}\over\!2}}
\nc{\intlv}{\int\limits_{-L}^{+L}\!\!{{{\rm d}v^-}\over\!2}}
\nc{\intv}{\int\limits_{-V}^{}\!\!{\rm d}^3\ulix}
\nc{\intvy}{\int\limits_{-V}^{}\!\!{\rm d}^3\uliy}
\nc{\zmint}{\int\limits_{-L}^{+L}\!\!{{{\rm d}x^-}\over{\!2L}}}
\nc{\zminty}{\int\limits_{-L}^{+L}\!\!{{{\rm d}y^-}\over{\!2L}}}
\nc{\intp}{\int\limits_{0}^{+\infty}\!\!{{{\rm d}p^+}\over\!4\gp}}
\nc{\intpp}{\int\limits_{0}^{+\infty}\!\!{\rm d}p^+}
\nc{\intlf}{\int\limits_{0}^{\infty}}
\nc{\intk}{\int\limits_{0}^{+\infty}\!\!dk^+}
\nc{\ink}{\int\limits_{0}^{+\infty}}
\nc{\inpm}{\int\limits_{-\infty}^{+\infty}}
\nc{\intkp}{\int\limits_{0}^{+\infty}\!\!\frac{dk^+}{\sqrt{4\gp k^+}}} 
\nc{\intkm}{\int\limits_{0}^{+\infty}\!\!\frac{dk^-}{\sqrt{4\gp k^-}}} 
\nc{\inp}{\int\limits_{0}^{\infty}\!\!{{\rm d}p^+}}
\nc{\inq}{\int\limits_{0}^{\infty}\!\!{{\rm d}q^+}}
\nc{\inpmi}{\int\limits_{0}^{\infty}\!\!{{\rm d}p^-}}
\nc{\inqmi}{\int\limits_{0}^{\infty}\!\!{{\rm d}q^-}}
\nc{\inpp}{\int\limits_{0}^{\infty}\!\!{{{\rm d}p^+}\over{\!2p^+\sqrt{2\gp}}}}
\nc{\inqq}{\int\limits_{0}^{\infty}\!\!{{{\rm d}q^+}\over{\!2q^+\sqrt{2\gp}}}}
\nc{\insl}{\int\limits_{-L}^{+L}\!\!{\rm d}x}
\nc{\intex}{\int\limits_{-\infty}^{+\infty}\!\!{\rm d}x^1}
\nc{\intey}{\int\limits_{-\infty}^{+\infty}\!\!{\rm d}y^1}
\nc{\intep}{\int\limits_{-\infty}^{+\infty}\!\!{\rm d}p^1}
\nc{\inti}{\int\limits_{-\infty}^{+\infty}}
\nc{\inteq}{\int\limits_{-\infty}^{+\infty}\!\!{\rm d}q^1}
\nc{\intek}{\int\limits_{-\infty}^{+\infty}\!\!{\rm d}k^1}
\nc{\intel}{\int\limits_{-\infty}^{+\infty}\!\!{\rm d}l^1}
\nc{\inty}{\int\limits_{-\infty}^{+\infty}\!\!{\rm d}y^-}
\nc{\intz}{\int\limits_{-\infty}^{+\infty}\!\!{\rm d}z^-} 
\nc{\incp}{\int d^2p} 
\nc{\incq}{\int d^2q} 
\def\beq{\begin{equation}}
\def\eeq{\end{equation}}
\def\bea{\begin{eqnarray}}
\def\eea{\end{eqnarray}}
\nc{\intx}{\int\limits_{-\infty}^{+\infty}\!\!\frac{{\rm d}x^-}{2}}
\nc{\intgix}{\int\limits_{-\infty}^{+\infty}\!\!{{{\rm d}x^-}\over\!2}}
\nc{\intgiy}{\int\limits_{-\infty}^{\infty}\!\!{{{\rm d}y^-}\over\!2}}

\begin{document}
\title{Vacuum loops in light-front field theory}  
\author{L$\!\!$'ubom\'{\i}r Martinovi\u{c}}  
\email{fyziluma@savba.sk}
\affiliation{Institute of Physics, Slovak Academy of Sciences \\
D\'ubravsk\'a cesta 9, 845 11 Bratislava, Slovakia \\ 
and \\
BLTP JINR, 141 980 Dubna, Russia}  
\author{Alexander Dorokhov} 
\email{dorokhov@theor.jinr.ru}    
\affiliation{Bogoliubov Laboratory of Theoretical Physics, JINR, 
141 980 Dubna, Russian Federation}  

\begin{abstract}
We demonstrate that vacuum diagrams in "native" light front (LF) field theory 
are non-zero, in spite of simple kinematical counter-arguments (positivity 
and conservation of the LF momentum $p^+$, absence of Fourier zero mode).  
Using the light-front Hamiltonian  
(time-ordered) perturbation theory, the vacuum amplitudes in self-interacting 
scalar $\lambda\phi^3(1+1)$ and $\lambda\phi^4(1+1)$ models are obtained as 
$p=0$ limit of the associated self-energy diagrams, where $p$ is the external 
momentum. They behave as $C\lambda^2\mu^{-2}$ in D=2, with $\mu$ being the 
scalar-field mass, or diverge in D=4, in agreement with the usual "equal-time" 
form of field theory, and with the same value of the constant $C$. The simplest 
vacuum diagram with two internal lines is analyzed in detail displaying the 
subtle role of the small $k^+$ region and its connection to the $p=0$ limit.    
However, the vacuum bubbles in the genuine light-front field theory are 
nonvanishing not due to the Fourier mode carrying LF momentum $k^+=0$ (as is 
the case in the LF evaluation of the covariant Feynman diagrams), in full 
accord with the observation that the LF perturbation theory formula breaks 
down in the exact zero-mode case. This is made explicit using the DLCQ method 
- the discretized (finite-volume) version of the theory, where the light-front 
zero modes are manifestly absent, but the vacuum amplitudes still converge to 
their continuum-theory values with the increasing "harmonic resolution" K. \\
\end{abstract}
\maketitle


\section{Introduction}  
Quantum field theory (QFT) formulated in terms of light-front (LF) variables 
\cite{Dir,KS,ChMa,LKS,BPP} has a few unusual features which are sometimes 
interpreted as indicating its inconsistency or incompleteness \cite{McCR,Burk}. 
One problem 
appeared to be paradoxically related to the most celebrated property of the 
LF quantization, namely vacuum simplicity. As is well known, positivity of 
the LF momentum $p^+$ together with its conservation implies that the ground 
state of any interacting model cannot contain quanta carrying $p^+\neq 0$. Only 
a tiny subset of all field modes, namely those carrying $p^+=0$ - the dynamical 
LF zero modes (ZMs) - can contribute. 
In addition, some modes (like the scalar-field zero mode) which appear as 
dynamical ones in the conventional ("space-like", SL for short) theory become 
constrained, that is, non-dynamical, in the LF form of the interacting theory 
(or vanish in the free theory), and hence cannot contribute to vacuum 
processes directly. The subtle role of the constrained zero modes   
has been known since the work of Maskawa and Yamawaki \cite{MY}, who  
analyzed the self-interacting scalar models using the discretized light-cone 
quantization (DLCQ), that is a finite-volume 
formulation with fields satisfying (anti)periodic boundary conditions in space 
variables. A natural question is if the LF formalism with its "trivial" vacuum 
is then capable to describe phenomena, related to nontrivial vacuum structure, 
like spontaneous symmetry breaking. More generally, is the LF form of QFT  
equivalent to the usual SL one, can it reproduce the known results or even 
predict some truly new insights?  

The equivalence issue has been realized and studied by Chang and Ma and by 
T.-M. Yan already in the pioneering LF papers \cite{ChMa,Yan}, including the 
vacuum problem at the level of the perturbation theory (i.e., the vacuum 
bubbles). Here one should distinguish two methods. In the first method, one 
starts from the covariant Feynman amplitudes and rewrites the corresponding 
integrals in terms of LF variables $k^{\pm}=k^0 \pm k^3$. The delicate step is 
to integrate over the $k^-$ variable (using the contour integration and the 
residue theorem) since the propagators (in two dimensions)  
behave as $(k^+k^- - m^2 +i\epsilon)^{-1}$ instead of $(k_0^2-k_1^2 - 
m^2 +i\epsilon) ^{-1}$, so there arises the convergence issue. Moreover, 
vacuum loop diagrams (diagrams with no external lines) seemed to vanish 
altogether when calculated in terms of LF variables. However, when the naive 
evaluation was replaced by a careful mathematical treatment \cite{ChMa,Yan}, 
these vacuum amplitudes were found to be nonzero. The missing contribution 
was identified as coming from the $k^+=0$ field mode due to the delta 
function $\delta(k^+)$ resulting from the $k^-$ integration \cite{ChMa,Yan}. 
Later on, this kind of zero modes turned out to be equally important in the case   
of non-vacuum LF Feynman amplitudes of various precesses related to 
LF wavefuctions, electroweak currents and formfactors \cite{melost,brohw,bakji}.    

The $k^+=0$ modes have been also found to give non-negligible contributions in 
model studies of pair creation \cite{tomrs,ild}, in the LF analysis of the  
effective potential \cite{hein1,hein2} and in the LF path integral formulation 
\cite{hein2}. It was shown in \cite{tanigy1,tanigy2} that when the ZM was 
missed in the discretized evaluation of the Feynman integrals expressed 
in terms of the LF variables (summation in $k^+$, integration in $k^-$), the 
forward scattering amplitude was not correctly reproduced in the continuum limit.   
This violation of Lorentz invariance is actually an artifact of the method used,  
however. In the genuine (non-manifestly covariant) DLCQ perturbation theory, 
where $k^-$ integration is not present (as is not the $k^+=0$ Fourier mode \cite{MY}), 
the limit of the forward-scattering amplitude is reached smoothly to an 
arbitrary precision with the increased value of the resolution parameter $K$ 
(i.e., in the continuum limit) \cite{lmhava}. 
 
The recent study \cite{mannh} of the "vacuum-sector diagrams", concentrating 
on the tadpole graph of the four-dimensional scalar field and claiming failure 
of the LF Hamiltonian method due to the intrinsic impossibility to capture 
the contribution of an arch at infinity, deserves a separate discussion 
\cite{lmannh}.
   
So far we have mostly mentioned the applications of the first method of 
computation of LF amplitudes, based on Feynman diagram technique. The second 
method is the LF perturbation theory (LFPT), which is not manifestly covariant 
and uses energy denominators instead of Feynman propagators. The integration 
over the $k^-$ variable is not present by construction.   
The same simple kinematical argument (positivity and conservation of $p^+$)  
suggests that vacuum bubbles are zero in LFPT. A vacuum diagram can be 
non-vanishing only if all internal lines carry $k^+=0$, but the rules of 
LFPT give an ill-defined result in this case \cite{Yan} (see the formula 
(\ref{Yvac}) below.) Thus, it has not been clear for a long time how the  
vacuum amplitudes can be correctly evaluated in the LF perturbation theory. 
They have often been considered as vanishing. One implication was that the 
cosmological constant problem could be solved in this way \cite{bros}. 
If true, this would mean that the conventional SL form and the LF form of 
the relativistic dynamics are not equivalent, since the vacuum loops  
diverge (logarithmically or quadratically) in the SL theory in D=3+1.    

Recently, J. Collins drew attention to this contradiction \cite{coll}. He 
used an example with the simplest vacuum diagram (two propagators) to 
identify the mathematical inconsistency present in the LF computation of the 
corresponding integrals. He also indicated an alternative approach, based 
on analyticity, to calculate the LF vacuum amplitudes as the limit  
of vanishing external momentum of the associated self-energy diagrams.  

In the present letter, we extend this argument to the vacuum bubbles of 
two-dimensional $\phi^3$ and $\phi^4$ models. We show that the self-energy 
diagrams, when represented in the LF perturbation theory, reduce for 
external momentum $p=0$ to nonzero limits equal to the values of vacuum 
bubbles known from the conventional Feynman diagrams. For example, the value  
of the vacuum loop of the $\lambda\phi^3$ model is given by the $p=0$ limit of 
the two-loop self-energy of the $\lambda\phi^4$ model. Moreover, 
the non-vanishing of LF vacuum loops is not due to the Fourier mode carrying  
$k^+=0$: same results as in the continuum theory are obtained 
in the DLCQ framework, namely in the finite-volume formulation with 
(anti)periodic boundary conditions, where the zero mode is manifestly 
absent. In this approach, the spectrum of field modes is discrete, 
enumerated by non-negative integers $n, 1\leq n\leq K-1$, where $K$ is called 
the "harmonic resolution". We show that although the Fourier modes with $n=0$ 
are not present, the corresponding summations (which replace integrals of 
the continuum theory) still converge to the same values of the vacuum 
amplitudes for increasing values of $K$. This adds further evidence to 
the attitude that the LF form of QFT is fully equivalent to the usual SL 
one. However, the mechanisms and mathematical appearance may be different, 
and the LF theory requires fresh and independent thinking.             

\section{Vacuum amplitudes in the conventional and light-front theory}
In the language of perturbation theory, the non-trivial structure of the 
ground state in relativistic QFT manifests itself by vacuum-polarization 
amplitudes (diagrams). In the case of the two-dimensional self-interacting 
$\lambda\phi^3$ and $\lambda\phi^4$ theories, these diagrams are shown in 
Fig. 1. Such a process of creation of two or more particles 
from the vacuum does not violate the momentum conservation in the SL theory 
since $p^1$ can 
acquire both the positive and negative values which add to zero (momenum 
conservation). Once again, this seems impossible in the LF case where $p^+$ 
values carried by the two quanta are strictly positive (or vanishing if there 
exist dynamical LF zero modes in the considered model).  

Let us remind briefly how the vacuum amplitudes are calculated in the SL form 
with the example of the $\phi^3$ bubble (see Fig. 1(b)). The corresponding 
Feynman rules lead to the double two-dimensional integral expression   
\bea 
&&\!\!\!\!\!\!\!\!\!\!\!\!\!\!\!\!\!\!\!\!V_3(\mu)= N_4\gl^2\int d^2k_1\int 
d^2k_2~G(k_1)G(k_2) G(k_1+k_2),~ \label{vb3} \\
&&\!\!\!\!\!\!\!\!\!\!\!\!\!\!\!\!\!\!\!\!G(k)=\frac{i}{k^2-\mu^2+i\epsilon}, 
~~~~N_4=\frac{1}{3!}\frac{1}{i(2\pi)^4}.  
\label{nvb3}
\eea 
The coefficient $1/3!$ is the symmetry factor. The double integral (\ref{vb3}) 
can be 
evaluated in a few ways: by using the Feynman parameters \cite{fey}, 
by means of $\ga$-representation \cite{smi} or via more sophisticated 
mathematical methods (Mellin-Barnes representation for powers of massive 
propagators \cite{datau}). All of them yield the same result
\beq
V_3(\mu)= -\frac{i\gl^2}{\mu^2}\pi^2N_4C,~~C=2.343908...,   
\label{vacres}
\eeq
where the constant $C$ has a particular representation in each of the 
computational method. For example, the first method requires to combine the 
propagators into one denominator by means of the auxiliary integrals in 
terms of the Feynman parameters $x_i$, then to go over to Euclidean space  
and calculate the integrals in $k_1$ and $k_2$ variables. The result is the  
double-integral representation  
\bea 
&&\!\!\!C=\int\limits_0^1 d x_1\int\limits_0^{1-x_1}d x_2~\frac{1}{D(x_1,x_2)}, 
\\
&&\!\!\!D(x_1,x_2)=x_1(1-x_1) + x_2(1-x_2) - x_1 x_2. \nonumber    
\label{xxx}  
\eea 
The first integral can be calculated analytically in terms of $arc tan$ 
function and square roots of polynomials, the numerical evaluation of the 
second integral then yields the above value of C. 

If we consider the self-energy diagram instead of the vacuum bubble 
($G(k_1+k_2)$ replaced by $G(p-k_1-k_2)$ in Eq.(\ref{vb3})), the analogous 
calculation yields 
\bea 
&&\!\!\!\!\!\gS_4(p^2)=N_4\int\limits_0^1 dx_1\int\limits_0^{1-x_1} dx_2
\frac{\gl^2}{A(x_1,x_2)p^2-D(x_1,x_2)\mu^2},\nonumber \\  
&&\!\!\!\!\!A(x_1,x_2)=x_1x_2(1-x_1-x_2). 
\eea 
Our goal now is to demonstrate that the result (\ref{vacres}) can be obtained  
also in the LF perturbation theory, contrary to   
the general belief. Let us recall that in the LFPT \cite{Weinb,ChMa,KS,LB}, no  
$T$-product is introduced. The integrals over the LF time variables are  
explicitly performed in the iterative solution of the equation for the 
$S$-matrix in the interaction representation. As the result, the particles  
in the intermediate states are on mass-shell with the usual Feynman propagators 
replaced by the LF energy denominators. The loss of manifest covariance is of 
no harm and the advantage is that one does not need to perform extra 
integration over the LF energy in the perturbative amplitudes. However, when 
the LFPT rules are applied 
to vacuum diagrams, the result is ill-defined \cite{Yan},  
as the delta function requires all three momenta to be equal to zero:  
\beq
\tilde{V} \sim \intlf \frac{dk^+_1}{k^+_1}\intlf \frac{dk^+_2}{k^+_2}\intlf  
\frac{dk^+_3}{k^+_3}\frac{\gd(k^+_1+k^+_2+k^+_3)}{(-\mu^2)[
\frac{1}{k^+_1}+\frac{1}{k^+_2}+\frac{1}{k^+_3}]}.     
\label{Yvac}
\eeq

\begin{figure*}
$$\includegraphics[width=0.7\linewidth]{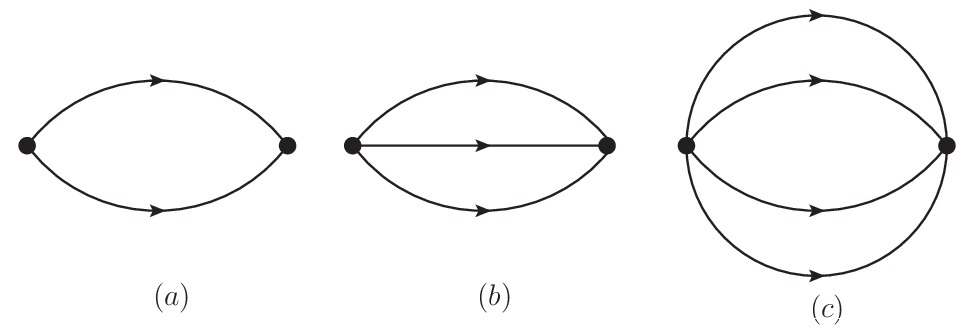}$$
\caption{Vacuum diagrams (a) $\gS_3(0)\equiv V_2(\mu)$ (equal to the Green 
function of $\phi^2(x)\phi^2(0)$ composite operator in free theory \cite{coll}), 
(b) $\gS_4(0)\equiv V_3(\mu)$ for the $\phi^3$ model and 
(c) for $\phi^4$ model (not discussed in the main text). 
The arrows indicate the momentum flow.}
\end{figure*}

It is actually not very difficult to resolve this inconsistency. One simply  
has to start from the graph with nonvanishing external momentum $p$ and to 
compute the $p=0$ limit in the corresponding integrals. The assumed  
analyticity in $p$ guarantees existence of this limit which should be equal 
to the vacuum amplitude. 
In this way, the expression 
(\ref{Yvac}) is first replaced by the self-energy \cite{foot}    
\bea 
&&\!\!\!\!\!\!\!\!\!\!\!\!\!\!\!\!\!\!\!
\gS_4(p^2)=\tilde{N}_4\gl^2\int\limits_0^{p^+}\frac{dk_1^+}{k_1^+}\int
\limits_0^{p^+-k_1^+}\frac{dk_2^+}{k_2^+(p^+-k_1^+-k_2^+)} \nonumber \\
&&\!\!\!\!\!\!\!\!\!\!\!\!\!\!\!\!\!\times\frac{1}{p^--\frac{\mu^2}{k_1^+}-
\frac{\mu^2}{k_2^+}- \frac{\mu^2}{p^+-k_1^+-k_2^+}+i\epsilon},~~
\tilde{N}_4=\frac{1}{3!}\frac{1}{(4\pi)^2}.   
\label{p3}
\eea 
Introducing the dimensionless variables 
$x=\frac{k_1^+}{p^+}$, $y=\frac{k_2^+}{p^+}$, $\gS_4(p)$ becomes
\beq
\gS_4(p^2)=\int\limits_0^1\frac{dx}{x}\int\limits_0^{1-x}\frac{dy}{y(1-x-y)}
\frac{\tilde{N}_4\gl^2}{p^2-\frac{\mu^2}{x}-\frac{\mu^2}{y}-\frac{\mu^2}
{1-x-y}}. 
\label{p2xy}
\eeq
Now we can set $p=0$ \cite{foota}. 
The integral over the variable $y$ can be performed explicitly, yielding  
\beq
F(x)=\frac{1}{\mu^2}\frac{4x}{\sqrt{3x^2-2x-1}}\arctan{\frac{1-x}
{\sqrt{3x^2-2x-1}}}.  
\label{firstin}
\eeq
The numerical computation of the second integral 
\beq
\gS_4(0)=\gl^2\tilde{N}_4\int\limits_0^1\frac{dx}{x}F(x) 
\label{secondi}
\eeq
gives
\beq
\gS_4(0)\equiv V_3(\mu)=-\frac{\gl^2}{\mu^2}\tilde{N}_4C,~~C=2.343908\dots,  
\label{lfc}
\eeq
in the complete agreement with the space-like result (\ref{vacres}).  
The overall situation is in fact very simple. Multiplying out the terms in the 
denominator of Eq.(\ref{p2xy}) for $p=0$, we find that the corresponding 
double integral is precisely equal to the representation of the constant $C$ in 
Eq.(4). The LF and SL schemes match at this point. The only difference 
is that in the SL theory one can start directly with the vacuum diagram while 
in the LF case one has to consider the associated self-energy diagram first.  
Its $p=0$ value can be taken after changing the LF momenta to relative 
variables which makes the integrand manifestly covariant, 
that is depending symmetrically on both $p^+$ and $p^-$. Obviously, 
$\gS_4(p^2)$ is also the same in the both schemes, the difference 
being that the LF scheme needs for that just two steps while the conventional 
Feynman procedure is considerably lengthier.  

\section{The simplest diagram with two internal lines}
The above result can be understood in more detail in a simpler case of   
the one-loop self-energy diagram in the $\lambda\phi^3$ theory (see Fig. 1(a) 
with two external lines attached). The corresponding Feynman amplitude is   
\beq
-i\gS_3(p^2)=\frac{1}{2}\frac{(-i\lambda)^2}{(2\pi)^2}\int d^2k~G(k)G(p-k),  
\label{f3l}
\eeq
The vacuum bubble ($p=0, \lambda=1$) rewritten in terms of the LF variables 
takes the form 
\beq
V_2(\mu)=\frac{i}{16\pi^2}\inpm dk^+\inpm dk^-\frac{1}{(k^+k^-- 
\mu^2+i\epsilon)^2}. 
\label{lf2l}
\eeq
To correctly evaluate the integral over $k^-$, one has to impose a cutoff 
$\Lambda$ \cite{Yan}, leading to ($\tilde{N}_3=-i/16\pi^2$)  
\beq   
V_2(\mu)=\tilde{N}_3\inpm\frac{dk^+}{k^+}\Big[\frac{1}{\Lambda k^+ - 
\mu^2+i\epsilon} - \frac{1}{-\Lambda k^+ - \mu^2 + i\epsilon}\Big].  
\nonumber
\label{tlint}
\eeq  
This integral can be equivalently rewritten as \cite{Yan}
\beq
V_2(\mu)=\tilde{N}_3\frac{\Lambda}{\mu^2}\inpm dk^+\Big[\frac{1}{\Lambda k^+ - 
\mu^2+i\epsilon} - \frac{1}{\Lambda k^+ + \mu^2 - i\epsilon}\Big].  
\nonumber
\label{yanid}
\eeq
For $\Lambda\rightarrow\infty$ we get  
\bea  
&&V_2(\mu)=\frac{\tilde{N}_3}{\mu^2}\inpm dk^+\Big[\frac{1}{k^++i\epsilon} - 
\frac{1}{k^+-i\epsilon}\Big] = \nonumber \\
&&=\frac{\tilde{N}_3}{\mu^2}(-2\pi i) \inpm dk^+
\delta(k^+)=-\frac{1}{8\pi}\frac{1}{\mu^2}. 
\label{Flf2}
\eea
This result coincides with the one obtained in the usual SL 
calculation as well as with the direct LF computation \cite{coll,lmh}. 
The same diagram, being the simplest one, sheds light upon the 
mechanism at work in the genuine LF case. The LFPT formula for the 
self-energy $\gS_3$ (\ref{f3l}) is ($\lambda=1$) 
\bea 
\!\!\!\!\!\!\!\!\!\!\!\!\!
\gS_3(p)=\frac{1}{8\pi}\int\limits_0^{p^+} \frac{dk^+}
{k^+(p^+-k^+)}\frac{1}{p^--\frac{\mu^2}{k^+}-\frac{\mu^2}{p^+-k^+}+ i\epsilon}. 
\label{lfpk}
\eea 
Going over to the variable $x=k^+/p^+$, one finds
\beq 
\gS_3(p)=\frac{1}{8\pi}\int\limits_0^1 \frac{dx}
{p^2x(1-x) - \mu^2 + i\epsilon}. 
\label{xin}
\eeq
For $p=0$ one indeed easily reproduces $V_2(\mu)$ of (\ref{Flf2}) (see Fig. 1(a)). 
Alternatively, we can work directly with the form (\ref{lfpk}).  
Taking $p^+=p^-=\eta$ for simplicity, we have 
\beq   
\gS_3(\eta)=\frac{1}{8\pi}\int\limits_0^{\eta} \frac{dk^+}{k^+
(\eta-k^+)}\frac{1}{\eta-\frac{\mu^2}{k^+}-\frac{\mu^2}{\eta-k^+}+ i\epsilon}. 
\label{lfpk1}
\eeq
The integral can be evaluated exactly with the result
\bea 
\gS_3(\eta)&=&-\frac{1}{4\pi}\big(G(\eta)-G(0)\big), \nonumber \\
G(k)&=&\frac{\arctan
\Big(\frac{2k-\eta}{\sqrt{4\mu^2-k^2}}\Big)}{\eta\sqrt{4\mu^2-\eta^2}}.
\label{seta}
\eea 
The expansion for infinitesimal $\eta$ gives 
\beq
\gS_3(\eta)=-\frac{1}{8\pi}\frac{1}{\mu^2}\Big[1+\frac{\eta^2}{4\mu^2} + 
{\cal O}(\eta^4)\Big].
\label{infeta}
\eeq
One can see that the correct result is recovered for $\eta=0$. The technical  
reason is simple: the integrand in (\ref{lfpk1}) is $\eta^{-1}[k^+(\eta-k^+)
-\mu^2]^{-1}$. For very small $\eta$ the expression in the brackets has 
almost a constant value very close to $(-\mu^2)$ at the interval $(0,\eta)$, 
while the diverging $\eta^{-1}$ factor is canceled by the length $\eta$ of 
the integration domain. However, setting $\eta=0$ from very beginning as in 
the formula (\ref{Yvac}) yields the wrong (ill-defined) result. Note also 
that the change of variables $x=k^+/\eta$ transforms the integral (\ref{lfpk1}) 
to the form (\ref{xin}) with $p^2$ replaced by $\eta^2$, displaying clearly  
the relation between the small $k^+$ region and the $p \rightarrow 0$ limit.  

The above calculation can be confirmed in yet another way, using the 
limiting approach instead of Yan's cutoff in the Feynman amplitude 
(\ref{lf2l}). In the self-energy (\ref{f3l}) rewritten in terms of the LF 
variables,
\bea 
&&\!\!\!\!\!\!\Sigma_3(p^2) = \frac{i}{16\pi^2} \inpm dk^+ \inpm dk^-\frac{1}
{(k^+k^-- \mu^2+ i\epsilon)}
\nonumber \\ 
&&~~~~~~~~\times \frac{1}{\big((p^+-k^+)(p^- -k^-) -\mu^2 +i\epsilon
\big)},  
\label{alty}
\eea 
integration over $k^-$ can be performed using the method of residue, 
because with $p\neq 0$ the poles are at a finite distance from the 
origin. The result exactly matches the LFPT amplitude (\ref{lfpk}) and hence 
its value at $p=0$ obviously yields the correct value of the vacuum amplitude. 
 
\section{Calculations in a finite volume: the DLCQ method} 
It is remarkable that the same result for the LF vacuum bubble is obtained 
in the discretized (finite-volume) treatment with (anti)periodic boundary 
conditions (BC). In both cases, the field mode carrying $k^+=0$ is manifestly 
absent. The field expansion at $\xpl=0$ is 
\beq  
\phi(0,\xmin)=\frac{1}{\sqrt{2L}}\sum_{n}^{\infty}\frac{1}{\sqrt{k^+_n}}
\big[a_n e^{-ik^+_n\xmin} + a^\dagger_n e^{ik^+_n\xmin}\big], 
\label{dphi}
\eeq
where $k^+_n = 2\gp n/L$ and $L$ is the box lenght. The 
index $n$ runs over half-integers for antiperiodic BC, so there is no  
zero mode by construction, and over integers for periodic BC. In this case 
the term with $n=0$ is excluded because the field equation in the ZM sector 
$\mu^2\phi_0 = 0$ requires the field mode $\phi_0\equiv \phi(k^+=0)$ to vanish 
for $\mu\neq 0$. 

The DLCQ analog of the $\gS_3(p)$ amplitude is 
\bea
&&\!\!\!\!\!\!\!\!\!\!\!\!\!\!\!\!\!\!\gS_3(p)={\cal N}_3 \sum_{k^+}^{p^+}
\frac{1}{k^+ (p^+-k^+)} \frac{1}{\big[p^- -\frac{\mu^2}{k^+} - 
\frac{\mu^2}{p^+ - k^+ }\big]}, 
\label{gs3p} \\
&&\!\!\!\!\!\!\!\!\!\!\!\!\!\!\!\!\!\!p^+=\frac{2\gp}{L}K,~k^+\equiv 
k^+_n=\frac{2\gp}{L}n,~ n=1,2,\dots,K-1. 
\label{disc1}
\eea
${\cal N}_3$ is the normalization constant. For $p=0$, choosing periodic BC, 
$\gS_3(0)=V_2(\mu)$ becomes
\bea  
&&V_2(\mu)=-\frac{1}{8\pi\mu^2}S, \nonumber \\
&&S=\sum_{n=1}^{K-1} \frac{1}{n(K-n)} 
\frac{1}{\big[\frac{1}{n}+\frac{1}{K-n}\big]} = \frac{K-1}{K}.
\label{vsum}
\eea 
For $K \rightarrow \infty$, $S$ obviously converges to the continuum value 
$1$. The same result is obtained for the antiperiodic BC \cite{lmd}.  
In Table I, the smooth approach of the self-energy value to the vacuum-loop 
value as $p\rightarrow 0$ is shown.  
\begin{table}
\caption{Smooth approach of the one-loop self-energy amplitude $\gS_3(p)$ of 
the $\gl\phi^3$ model from Eq.(\ref{gs3p}) to its (rescaled) value $S=1-K^{-1}$ 
at $p=0$ ($K$=512).} 
\vspace{3mm}
\begin{tabular}{||c|c|c|c|c|c|c||}
\hline \hline
$p^2/\mu^2$ & $10^{-2}$ & $10^{-3}$ & $10^{-4}$ & $10^{-5}$ & $10^{-6}$ & 0 \\
\hline
$S$ & 1.50902 & 1.09320 & 1.00667 & 0.99890 & 0.99813 & 0.99805  \\
\hline \hline
\end{tabular}
\label{table:pC}
\end{table} 
The vacuum bubble with three internal lines can be computed in the same way. 
The discrete version of the self-energy (\ref{p2xy}) is  
\bea 
&&\!\!\!\!\!\!\!\!\!\!\!\!\!\!\gS_4(p)=\gl^2 {\cal N}_4 \sum_{q^+}^{p^+}
\sum_{k^+}^{p^+-q^+} \frac{1}{k^+ q^+ (p^+-k^+ - q^+)} \nonumber \\
&&~~~~~~~~~~~~\times \frac{1}{\big[p^- -\frac{\mu^2}{k^+} -\frac{\mu^2}{q^+} 
-\frac{\mu^2}{p^+ - k^+ - q^+}\big]},\\
&&\!\!\!\!\!\!\!\!\!\!\!p^+=\frac{2\gp}{L}K,~k^+\equiv k^+_n=\frac{2\gp}{L}n,~
q^+ \equiv q^+_m=\frac{2\gp}{L}m.
\label{disc2}
\eea
${\cal N}_4$ is a normalization constant. For $p=0$, choosing 
periodic BC, $\gS_4$ becomes
\bea 
&&\!\!\!\!\!\!\gS_4(0)=V_3(\mu^2)=-\frac{\gl^2}{\mu^2}\tilde{N}_4\sum_{m=1}^{K-2}
\frac{1}{m} \sum_{n=1}^{K-m-1}\frac{1}{n(K-m-n)} \nonumber \\
&&~~~~~~~~~~~~~~~~~~
\times\frac{1}{\big[\frac{1}{m}+\frac{1}{n}+\frac{1}{K-m-n}\big]}.
\label{disc3}
\eea
Numerical values of the double sum in $\gS_4(0)$ for increasing values of $K$ 
are shown in Table II. They readily 
approach the continuum value (\ref{lfc}) of the constant $C$. 
\begin{table}[t]
\caption{Convergence of the vacuum-loop constant $C$ from Eq.(\ref{lfc}) 
represented by the double sum of Eq.(\ref{disc3}) to its continuum value 
$C=2.343908...$ with the resolution $K$.}
\vspace{3mm}
\begin{tabular}{||c|c|c|c|c|c|c|c||}
\hline \hline
$K$ & 32 & 64 & 128 & 256 & 512 & 1024 & 2048 \\
\hline
$C$ & 1.921 & 2.099 & 2.205 & 2.266 & 2.301 & 2.320 & 2.331 \\
\hline \hline
\end{tabular}
\label{table:KC}
\end{table} 
The same pattern is valid also for the LF vacuum loops with four and more  
internal lines \cite{lmd}. This may be of interest  
for the LF solution of the sine-Gordon model \cite{Grif,Burk}.  
\section{Conclusions}
We have demonstrated that the vacuum amplitudes are non-zero in LF 
perturbation theory and match the known values. The calculations were 
performed within the self-interacting scalar models in two dimensions. 
Generalization of these results to higher dimensions where the vacuum loops 
diverge, is straightforward. One can also consider vacuum diagrams with  
internal lines corresponding to fields with different masses 
(the Yukawa model, e.g.).

Our results have been obtained first in the usual continuum form of field theory. 
Switching to the discretized (finite-volume) formulation, the values of vacuum 
amplitudes were shown to be reproduced for the resolution parameter $K$ tending 
to infinity (i.e., in the infinite-volume limit). As the scalar zero modes are 
manifestly absent in the finite-volume treatment with (anti)periodic boundary 
conditions, the non-vanishing of vacuum bubbles is solely due to non-zero field 
modes. However, to see this, one has to start from the associated diagram with 
non-zero incoming momentum $p$ and take the limit $p\rightarrow 0$ at the end 
of the calculation. Starting with $p=0$ leads to mathematically ill-defined 
amplitudes like Eq.(\ref{Yvac}). Let us mention here that the limiting approach 
of the present work was used previously in the context of the LF scattering 
amplitudes \cite{lmhava} computed in DLCQ. We note also that the LF momentum 
is conserved in the diagram vertices for arbitrarily small value of $p$ and 
hence in the limiting sense also for $p=0$. This observation is reinforced by 
the fact that the LF self-energy diagrams when re-expressed in terms of relative 
variables for $p=0$ exactly match the covariant vacuum amplitudes expressed in 
terms of Feynman parameters. In view of this it is not too surprising that the 
DLCQ evaluation gives the correct answers, as it is just a discrete approximation 
to the corresponding integrals.   

On the other hand, as emphasized in \cite{coll}, non-vanishing of vacuum bubbles 
in the genuine LF theory does not spoil its fundamental property, namely the 
possibility to construct a consistent Fock expansion for composite systems with 
well-defined coefficients - the LF wave functions. The crucial feature that 
distinguishes the LF formulation from the conventional SL one is that 
in the former, creation and annihilation operators can be consistently defined 
even in the interaction case, so that the vacuum-annihilation condition exists 
for an interacting theory in the Heisenberg picture \cite{LKS,yam98}. The ultimate 
reason for this is the positivity of the spectrum of a kinematical quantity - 
the LF momentum $P^+$, in addition to positivity of the energy $P^-$. 
  
In summary, while in the evaluation of the covariant vacuum Feynman diagrams 
in terms of the LF variables for the self-interacting scalar models the 
non-zero values of the vacuum amplitudes are due to the sharp $k^+=0$ contribution 
(with $k^+$ being the loop variable), the situation in the genuine LF theory 
is somewhat different. The Fourier mode carrying $k^+=0$ is not present in 
the LF perturbation theory but the correct results   
are nevertheless obtained if one uses a regularized approach, which adequatly 
treats the integration region close to $k^+=0$. Passing to the 
relative LF variables (which coincide with the Feynman parameters) maps this 
picture to the $p=0$ limit of the corresponding self-energy Feynman diagrams.   
 
\vspace{4mm}
\section{Acknowledgements} The authors are grateful to V. Karmanov for many useful 
discussions. LM thanks the Slovak Committee for the collaboration with JINR 
for a long-term support.


\begin{thebibliography}{99}
\bibitem{Dir} P.A.M. Dirac, Forms of relativistic dynamics, Rev. Mod. Phys. 21 
(1949) 392-399. 
\bibitem{KS} J. Kogut, D. E. Soper, Quantum electrodynamics in the infinite 
momentum frame, Phys. Rev. D1 (1970) 2901-2913. 
\bibitem{ChMa} S.-J. Chang, S. K.-Ma, Feynman rules and quantum electrodynamics 
at infinite momentum, Phys. Rev. 180 (1969) 1506-1513. 
\bibitem{LKS} H. Leutwyler, J. R. Klauder, L. Streit, Quantum field theory on 
lightlike slabs, Nuovo Cim. A66 (1970) 536-554. 
\bibitem{BPP} S. J. Brodsky, H.-C. Pauli, S. S. Pinsky, Quantum chromodynamics and 
other field theories on the light cone, Phys. Rept. 301  
(1998) 299-486. 
\bibitem{McCR} G. McCartor, D. G. Roberston, Light cone quantization of gauge 
fields, Z. Phys. C62 (1994) 349-356. 
\bibitem{Burk} M. Burkardt, Light front quantization of the sine-Gordon model, 
Phys. Rev. D47 (1993) 4628-4633. 
\bibitem{MY} T. Maskawa, K. Yamawaki, The Problem of $P^+=0$ Mode in the Null-Plane 
Field Theory and Dirac's Method of Quantization, Prog. Theor. Phys. 56 (1976) 
270-283. 
\bibitem{Yan} T.-M. Yan, Quantum field theories in the infinite momentum frame.4., 
Phys. Rev. D7 (1973) 1780-1800. 
\bibitem{melost} J.P.B.C. de Melo, J.H.O. Sales, T. Frederico, P.U. Sauer, 
Pairs in the light-front and covariance, Nucl.Phys. A631 (1998) 574c-579c. 
\bibitem{brohw} S. J. Brodsky, D. S. Hwang, Exact light cone wave function 
representation of matrix elements of electroweak currents, Nucl. Phys. B543 
(1998) 239-252. 
\bibitem{bakji} B.L.G. Bakker, H.-M. Choi, Ch.-R. Ji, The vector meson form-factor 
analysis in light front dynamics, Phys. Rev. D65 (2002) 116001. 
\bibitem{tomrs} T.N. Tomaras, N.C. Tsamis, R.P. Woodard, Pair creation and axial 
anomaly in light cone QED(2), JHEP 11 (2001) 008. 
\bibitem{ild} A. Ilderton, Localisation in worldline pair production and lightfront 
zero-modes, JHEP 09 (2014) 106.  
\bibitem{hein1} Th. Heinzl, The Light-Cone Effective Potential, 
arXiv:hep-th/0212202 [hep-th]. 
\bibitem{hein2} Th. Heinzl, Light cone zero modes revisited, 
arXiv:hep-th/0310165 [hep-th]. 
\bibitem{tanigy1} M. Taniguchi, S. Uehara, S. Yamada, K. Yamawaki, Does DLCQ 
S-matrix have a covariant continuum limit?, Mod. Phys. Lett. A 16 (2001) 2177-2185. 
\bibitem{tanigy2} M. Taniguchi, S. Uehara, S. Yamada, K. Yamawaki, Recovering 
Lorentz invariance of DLCQ, Arxiv:hep-th/0309240[hep-th].  
\bibitem{lmhava} A. Harindranath, L. Martinovic and J. P. Vary, Perturbative S 
matrix in discretized light cone quantization of two-dimensional $\phi^4$ theory, 
Phys. Lett. B 536 (2002) 250-258. 
\bibitem{mannh} P. D. Mannheim, P. Lowdon, S. J. Brodsky, Structure of light-front 
sector vacuum diagrams, Phys. Lett. B 797 (2019) 134916 
\bibitem{lmannh} L. Martinovic, talk at the Light Cone 2019 conference, Palaiseau, 
September 16-20 2019, and to be published. 
\bibitem{bros} S.J. Brodsky, R. Schrock, Condensates in quantum chromodynamics and 
the cosmological constant, Proc. Nat. Acad. Sci. 108 (2011) 45-50. 
\bibitem{coll} J. Collins, The non-triviality of the vacuum in light-front 
quantization: An elementary treatment, arXiv:1801.03960[hep-ph]. 
\bibitem{fey} R. P. Feynman, Space-time approach to quantum electrodynamics, 
Phys. Rev. 76 (1949) 769-789.
\bibitem{smi} V. A. Smirnov, {\it Analytic Tools for Feynman Integrals}, 
Springer Tracts in Mod. Phys., Springer-Verlag 2012.  
\bibitem{datau}A. I. Davydychev, J. B. Tausk, Two loop selfenergy diagrams with 
dfferent masses and the momentum expansion, Nucl. Phys. B397 (1993) 123-142. 
\bibitem{Weinb} S. Weinberg, Dynamics at infinite momentum, Phys. Rev. 159 (1966) 
879-883. 
\bibitem{LB} G. P. Lepage, S. J. Brodsky, Exclusive processes in quantum chromodynamics, 
Phys. Rev. D22 (1980) 2157. 
\bibitem{foot} It may appear that one is considering the 
self-energy of a model higher by one power in $\phi$, but it is  
sufficient to simply pretend that there is an incoming $p=(p^+,p^-)$ and 
write down the LF amplitude correspondingly.    
\bibitem{foota} One would prefer to set $p=0$ in the final formulas 
but it is rarely possible to compute multiple integrals analytically. 
\bibitem{lmd} L. Martinovic, A. Dorokhov, in preparation   
\bibitem{lmh} A. Harindranath, L. Martinovic, J.P. Vary, Compactification near and on 
the light front, Phys. Rev. D62 (2000) 105015. 
\bibitem{Grif} P. A.  Griffin, The sine-Gordon model and the small k+ 
region of light cone perturbation theory, Phys. Rev. D46 (1992) 3538-3543. 
\bibitem{yam98} K. Yamawaki, Zero-mode problem on the light front, 
arXiv:hep-th/9802037 [hep-th].  
\end{thebibliography}
\end{document}